\documentclass[12pt,a4paper,final]{iopart}
\usepackage{iopams}  
\usepackage{cite}
\usepackage{graphicx}
\usepackage[breaklinks=true,colorlinks=true,linkcolor=blue,urlcolor=blue,citecolor=blue]{hyperref}

\begin{document} 
\title[]{Two-flavour mixture of a few fermions of different mass in a one-dimensional harmonic trap}
 
\author{Daniel P{\c e}cak$^{1}$, Mariusz Gajda$^{1,2}$, and Tomasz Sowi\'nski$^{1,2}$}
\address{$^1$Institute of Physics of the Polish Academy of Sciences \\ Al. Lotnik\'ow 32/46, 02-668 Warsaw, Poland}
\address{$^2$Center for Theoretical Physics of the Polish Academy of Sciences \\ Al. Lotnik\'ow 32/46, 02-668 Warsaw, Poland}
\ead{pecak@ifpan.edu.pl}

\begin{abstract}
A system of two species of fermions of different mass confined in a one-dimensional harmonic trap is studied with an exact diagonalisation approach. It is shown that a mass difference between fermionic species induces a separation in the density of the lighter flavour independently of the number of particles. The mechanism behind the emergent separation is explained phenomenologically and confirmed by direct studies of the ground state of the system. Finally, it is shown that the separation driven by a mass difference, in contrast to the separation induced by a difference of populations, is robust to the interactions with thermal environment.

\end{abstract}

\section{Introduction}
Development of cooling and trapping techniques of a few quantum particles (bosons as well as fermions) have opened completely a new area of modern quantum engineering \cite{serwane2011deterministic,Blume2012,Thalhammer2006,Kohl2005,lewenstein2012ultracold}. In todays experiments, it is possible to rigorously control not only the strength of mutual interactions between atoms but also the precise number of quantum particles of different flavours confined in a trap of almost arbitrary shape \cite{serwane2011deterministic,PhysRevLett.100.053201,PhysRevLett.104.053202}. These possibilities are strongly related to the fundamental theoretical questions about the properties of a mesoscopic number of particles which up to now had not been answered in a satisfactory way. One of these questions is about the transition  from two-, three-body physics \cite{Schmitz2013Breathing,Blume2010,LiuHuDrummond2010Fermions3} that are quite well understood with standard approaches, to  macroscopic systems, that are well described with the help of sophisticated methods of condensed matter physics. A first attempt to give an experimental answer to this question was given in \cite{wenz2013few} where the emergence of the Fermi sea with increasing number of fermions was studied in a few-body system. In the attractive regime the effective appearance of two-body pairing as a function of the system's size was observed \cite{haller2009realization} and a theoretical description of this phenomenon was given \cite{Astrakharchik2008,Sowinski2014,Sowinski2015,DAmico2015Pairing}. Recently, many different results in the area of ultra-cold few-body systems were reported both in experimental \cite{zurn2012fermionization,paredes2004tonks,kinoshita2004observation,PhysRevLett.95.190406,PhysRevLett.100.010401,jo2009itinerant,PhysRevA.73.011402,PhysRevLett.92.120403} and theoretical \cite{Gersch1963,Olshanii1998,PhysRevLett.105.170403,PhysRevA.78.013613,daily2012thermodynamics,PhysRevA.90.013611,PhysRevA.90.023605,PhysRevLett.112.015301,bugnion2013ferromagnetic,PhysRevA.87.023605,loft2014variational,cui2014ground,Volosniev2013,Gharashi2013,DAmico2014,Yan2014,LiuHuDrummond2009Virial,KuhnFoerster2012,GarciaMarch2013,GarciaMarch2014Localization,GarciaMarch2014ThreeBosons,Eggert2011,Volosniev2014} works. 

In a recent paper \cite{ZinnerNJP}, a theoretical analysis of phase separation in a one-dimensional imbalanced mixture of two kinds of fermions was made. It was shown that strong interactions with a single impurity dipped in a mesoscopic Fermi sea lead to a separation of density in the bulk flavour. This separation is directly induced by the  imbalance in the number of fermions of the two types. In particular it can not be observed in a balanced system. However, it is known that in balanced systems consisting of flavours of different atomic mass, separation is also possible. For example, in the recent paper \cite{JasonHo}, density separation driven by a mass difference was shown in a system confined in a uniform trap. Since in experiments with ultra-cold fermionic mixtures it is possible to confine atoms with different masses \cite{PhysRevLett.100.053201,PhysRevLett.104.053202,PhysRevLett.100.010401,PhysRevA.73.011402,PhysRevA.78.013613}, therefore it is interesting to study the properties of such a system in detail. In this paper, we study the separation mechanism in a mixture of two-flavour fermionic atoms of different mass  confined in a one-dimensional harmonic trap. The whole analysis is performed with an exact diagonalisation of the many-body Hamiltonian.

At this point it is worth noting that Hubbard-like models with imbalanced mass of particles were also intensively studied \cite{takemori2012low,PhysRevA.87.053602}.

\section{The Model}
In this paper we study a two-flavour system of a few ultra-cold fermions confined in a one-dimensional harmonic trap of frequency $\omega$ and interacting via a short-range delta-like potential. The fermions of different flavours are fundamentally distinguishable and in principle may have different masses $m_\uparrow \neq m_\downarrow$. 
If one expresses all quantities in the natural units of the harmonic oscillator related to one of the flavours (in the following we choose~$\sigma=\downarrow$) then the Hamiltonian of the system in the second-quantized form reads:
\begin{equation}\label{Ham1}
 \hat{\cal H} = \sum_\sigma\int \mathrm{d}x\,\hat\Psi^\dagger_\sigma(x) H_\sigma \hat\Psi_\sigma(x) + g\int\mathrm{d}x\,
 \hat\Psi^\dagger_\uparrow(x)\hat\Psi^\dagger_\downarrow(x)\hat\Psi_\downarrow(x)\hat\Psi_\uparrow(x),
\end{equation}
where $\hat\Psi_\sigma(x)$ annihilates a fermion with spin $\sigma$ at a point $x$ and the dimensionless interaction constant $g=(m_\downarrow/ \hbar^3 \omega)^{1/2}g_\mathrm{1D}$ is related to the effective one-dimensional interaction $g_\mathrm{1D}$ between fermions of opposite spins. Experimentally this effective one-dimensional model is obtained by applying very strong harmonic confinement in two remaining perpendicular directions \cite{zurn2012fermionization}. Then it can be assumed that the dynamics in these directions is frozen and particles occupy only the ground state of the trap. A direct relation between effective interaction $g_\mathrm{1D}$ and the s-wave scattering length $a_s$ which can be tuned with Feshbach resonance was discussed with all the details in \cite{Olshanii1998}. The single-particle Hamiltonians $H_\sigma$ take into account the mass ratio $\alpha=m_\uparrow/m_\downarrow$ and have the following forms
\begin{equation}
H_\downarrow = \frac{1}{2}\frac{\mathrm{d}^2}{\mathrm{d}x^2} + \frac{1}{2}x^2, \qquad H_\uparrow = \frac{1}{2\alpha}\frac{\mathrm{d}^2}{\mathrm{d}x^2} + \frac{\alpha}{2}x^2. 
\end{equation}
It is worth noticing that in principle both flavours of fermions can be confined in traps with different frequencies. Here we assume that the experimental parameters are tuned such that the frequencies are equal. At the level of the single-particle Hamiltonian the only difference between flavours is related to the mass difference. This assumption has some consequences on the properties of the system. For example, the single-particle excitation energy is flavour-independent and it is equal to $\hbar\omega$. 

Since particles of different flavours are distinguishable, the field operators for opposite flavours fulfil standard commutation relations $\left[\hat{\Psi}_\uparrow(x),\hat{\Psi}^\dagger_\downarrow(x')\right]=\left[\hat{\Psi}_\uparrow(x),\hat{\Psi}_\downarrow(x')\right]=0$. However, the field operators for the same flavour obey the canonical fermionic anti-commutation relations $\left\{\hat{\Psi}_\sigma(x),\hat{\Psi}^\dagger_\sigma(x')\right\}=\delta(x-x')$ and $\left\{\hat{\Psi}_\sigma(x),\hat{\Psi}_\sigma(x')\right\}=0$. Obviously, the Hamiltonian (\ref{Ham1}) commutes separately with the number of fermions of a given flavour $\hat{N}_\sigma=\int\!\mathrm{d}x\,\hat\Psi^\dagger_\sigma(x)\hat\Psi_\sigma(x)$. Therefore, further analysis will always be performed under the assumption that the number of fermions in a given flavour is known and well defined experimentally. This assumption is reasonable since in ongoing experiments with few fermion systems the number of particles is controlled to a very high accuracy \cite{serwane2011deterministic,wenz2013few}.

\section{Method}
The properties of the system described by the Hamiltonian (\ref{Ham1}) are studied with the exact diagonalisation approach what was previously adopted for equal mass fermionic mixtures \cite{SowinskiGrass,Sowinski2014,Sowinski2015}, in both the repulsive and attractive case. First, we decompose the field operators $\hat{\Psi}_\sigma(x)$ onto the basis of the eigenfunctions of the appropriate single-particle Hamiltonian, i.e., $\hat{\Psi}_{\sigma} (x) = 
 \sum_i \phi_{i\sigma} (x) \hat{a}_{i \sigma}$. Note, that due to the mass difference, the basis functions of different flavours have different spatial spread and they are related by the scaling relation
 $\phi_{i\uparrow}(x) = \alpha^{1/4}\,\phi_{i\downarrow}(\sqrt{\alpha}\,x)$. In this language, the Hamiltonian (\ref{Ham1}) has the following form
\begin{equation} \label{Ham2}
\hat{\cal H} = \sum_\sigma \sum_i E_i\, \hat{n}_{i\sigma} + \sum_{ijkl}U_{ijkl}\,\hat{a}^\dagger_{i\uparrow}\hat{a}^\dagger_{j\downarrow}\hat{a}_{k\downarrow}\hat{a}_{l\uparrow},
\end{equation}
where $\hat{n}_{i\sigma}=\hat{a}^\dagger_{i\sigma}\hat{a}_{i\sigma}$.
The single-particle energy $E_i=(i+1/2)$ and the two-body interaction term $U_{ijkl}$ can be calculated directly 
\begin{equation} \label{IntU}
U_{ijkl} = g \int\mathrm{d}x\, \phi^*_{i\uparrow}(x)\phi^*_{j\downarrow}(x)\phi_{k\downarrow}(x)\phi_{l\uparrow}(x).
\end{equation}
Since the wave functions $\phi_{i\sigma}(x)$ form a complete set in the space of single-particle states, the Hamiltonian (\ref{Ham2}) is fully equivalent to the original Hamiltonian (\ref{Ham1}). In practice, we have to cut the decomposition of the field operators at some sufficiently large level index $N_{\mathrm{max}}$. For appropriately chosen cut-off, the results do not depend on its value. For example, for the parameters $N_{\uparrow}=3$, $N_{\downarrow}=1$, $\alpha=1$ and $g=10$ it is sufficient to take $N_{\mathrm{max}}=18$ single-particle eigenstates in the decomposition of the field operators. Then, the dimension of the many-body Hilbert space is $\mathrm{dim}(\boldsymbol{H})=14688$.

After obtaining the matrix form of the Hamiltonian (\ref{Ham1}) in the Fock basis for a given number of fermions and interaction strength $g$, and after subsequent diagonalisation, we find the $M$ lowest many-body eigenstates $|\mathtt{G}_i\rangle$ and corresponding eigenenergies ${\cal E}_i$ of the Hamiltonian.

\begin{figure}
 \centering
 \includegraphics[scale=1.2]{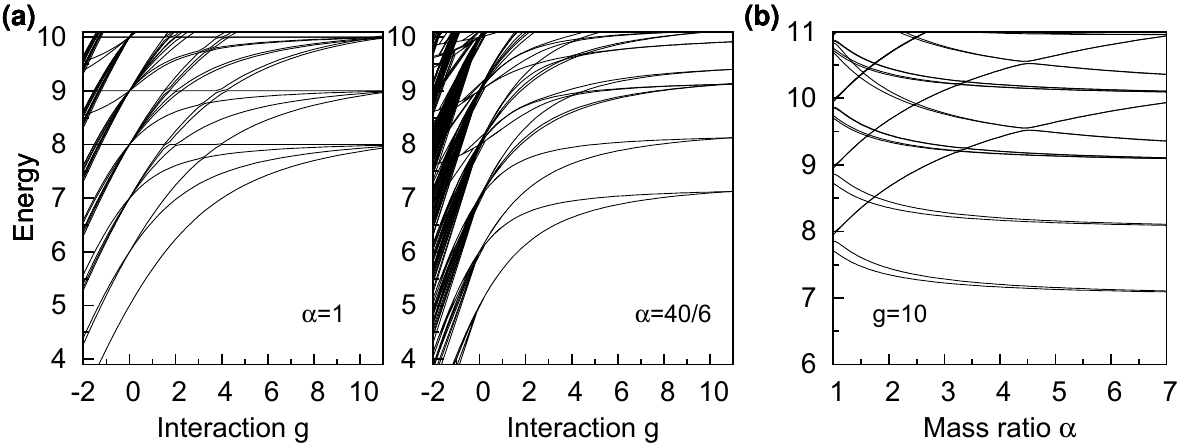}
 \caption{(a) Spectrum of the Hamiltonian (\ref{Ham1}) for $N_{\uparrow}=3$ and $N_{\downarrow}=1$ as a function of the interaction $g$. Depending on the ratio of  atomic masses $\alpha$ the spectra differ qualitatively for strong repulsion. For example, the degeneracy manifolds in the limit of $g\rightarrow \infty$ have different dimensions. (b) Spectrum of the Hamiltonian (\ref{Ham1}) for $N_{\uparrow}=3$ and $N_{\downarrow}=1$ and strong interaction $g=10$ as a function of the mass ratio $\alpha$. It is seen that for increasing $\alpha$, the energy of the two many-body eigenstates belonging to the ground-manifold rapidly split and the ground-manifold has only a two-fold degeneracy.  \label{fig1} }
\end{figure}

\section{Spectrum of the Hamiltonian}
The first significant consequence of the mass difference between flavours is visible in the spectrum of the Hamiltonian. In the standard situation, when $\alpha=1$, the Hamiltonian of the system is symmetric under exchange of the flavours. As a matter of fact, this symmetry, together with the Pauli exclusion principle, leads directly to the appearance of a degeneracy in the manifolds of many-body eigenstates of the Hamiltonian in the limit of infinite repulsion \cite{Girardeau,SowinskiGrass}. Each degenerate manifold is spanned by states with the same spatial distribution of particles but with different characters of the irreducible representation of the exchange-symmetry group mentioned above. For large but finite repulsions the spectrum of the Hamiltonian is not degenerate but it factorises into distant manifolds of quasi-degenerate eigenstates. The mechanism that lifts the degeneracy and its experimental consequences have been broadly discussed recently \cite{loft2014variational,SowinskiGrass}.

It is worth noticing that the case of $N_{\uparrow}=N_{\downarrow}=1$ is solvable analytically for any value of the mass ratio $\alpha$. The two-body problem is brought to a single-particle one by introducing centre of mass and relative distance coordinates. In these coordinates, the Hamiltonian of the system is decomposed into two single-particle problems for which analytical expressions for eigenenergies and eigenfunctions are known\cite{Busch1998}. Many properties of two particles confined in harmonic trap were recently studied in fermionic as well as in bosonic systems  \cite{zurn2012fermionization,Sowinski2010Dynamics,Esslinger2006,koscik2012quantum,koscik2012ground,PhysRevLett.108.115302}.

Whenever $\alpha\neq 1$, the exchange-symmetry is violated since the interaction part of the many-body Hamiltonian, in contrast to the single-particle one, is not symmetric under exchange of particles. This directly leads to different spectral properties of the Hamiltonian in the limit of large repulsion. To illustrate this mechanism, in Fig. \ref{fig1}a we compare spectra of the Hamiltonian (\ref{Ham1}) with $\alpha=1$ and $\alpha=40/6$ (corresponding to the mass ratio in the mixture of lithium and potassium atoms \cite{PhysRevLett.100.053201,PhysRevLett.104.053202,PhysRevLett.100.010401,PhysRevA.73.011402,PhysRevA.78.013613}) calculated for a fixed number of fermions $N_\uparrow=3$ and $N_\downarrow=1$ as a function of interaction $g$. As it can be seen, for  large interactions the degeneracy manifolds do split. Our numerical results suggest that this mechanism is present for different $N_\uparrow$ and $N_\downarrow$.
Unfortunately, due to the numerical complexity, our method enables us to study only those configurations for which $N_\uparrow+N_\downarrow\le 10$. However, for all of them the same mechanism is present if the mass ratio $\alpha$ is large enough. Therefore, we believe that the mechanism is very general and it is present for any number of particles.

It is quite natural that for different pairs of $N_\sigma$, the degeneracy may be lifted into sub-manifolds with different numbers of eigenstates. To give a better explanation of how this degeneracy is lifted we show in Fig. \ref{fig1}b the spectrum of the Hamiltonian (\ref{Ham1}) for $g=10$ as a function of the mass ratio $\alpha$. For $\alpha=1$ the ground-manifold is spanned by four many-body eigenstates. However, for a larger $\alpha$ the eigenenergy of the two states rapidly grows. Due to changes of the shapes of the corresponding single-particle wavefunctions $\phi_{i\sigma}$, the contact interaction between flavours becomes significant. It is seen that, the same mechanism of quasi-degeneracy  splitting  repeats also for higher manifolds.

\section{Properties of the ground-state}

\begin{figure}
 \centering
 \includegraphics[scale=1.2]{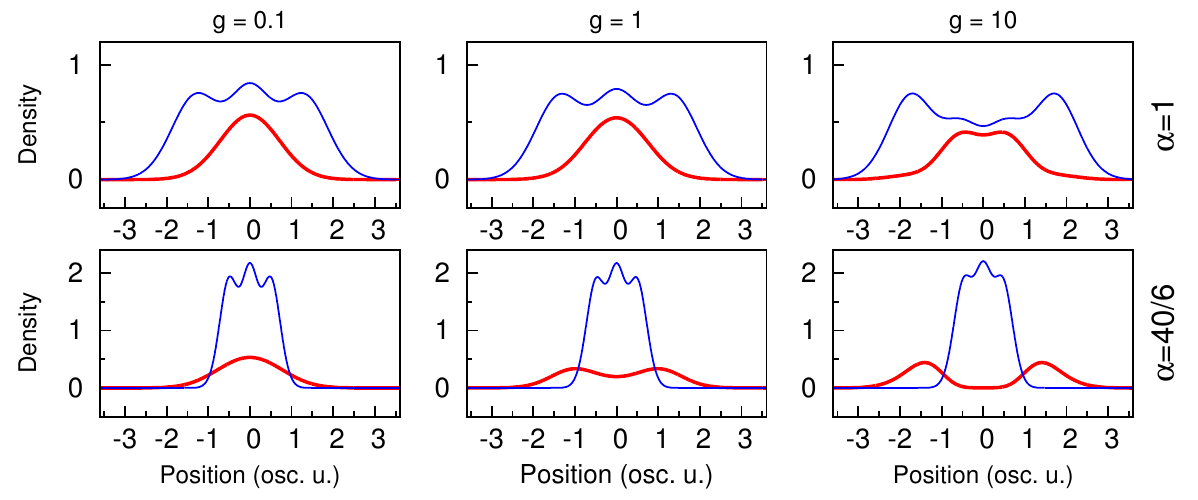}
\caption{Single-particle densities $\rho_\uparrow(x)$ (thin blue line, heavier flavour) and $\rho_\downarrow(x)$ (thick red line, lighter flavour) calculated in the ground-state of the system for different interactions $g$ and mass ratios $\alpha$ ($N_\uparrow=3$ and $N_\downarrow=1$). For weak interactions, the single-particle densities reproduce the density profile for ideal gasses in both species. When interaction is strong enough, a phase separation takes place in the system. For equal masses, the separation occurs in the species with the larger number of atoms (as predicted in \cite{ZinnerNJP}). On the contrary, for $\alpha=40/6$ the lighter species undergoes separation. We have checked that this behaviour is independent on the number of fermions in individual flavours (see main text and Fig. \ref{fig3}).  \label{Fig2} }
\end{figure}

The mass difference between fermionic species significantly changes not only the properties of the spectrum of the Hamiltonian but also some properties of many-body states. First, let us discuss changes of the many-body ground-state $|\mathtt{G}_0\rangle$. For $\alpha=1$ the properties of the ground-state were studied recently in \cite{ZinnerNJP}. It was shown that repulsive interactions between species lead to phase separation whenever the number of particles is imbalanced, i.e. $N_\downarrow\neq N_\uparrow$. In particular, the authors studied the interaction of a single impurity dipped in the Fermi sea of the remaining species ($N_\uparrow>1$ and $N_\downarrow=1$). They found that for strong interaction, the bulk flavour is pushed out of the centre of the trap and two spin domains precursors appear. The remaining impurity is located in the middle with an almost unchanged density profile. This observation can be visualized by plotting the single-particle densities defined as follows
\begin{equation} \label{DensProf}
\rho_\sigma(x) = \langle \mathtt{G}_0|\hat\Psi_\sigma^\dagger(x)\hat\Psi_\sigma(x)|\mathtt{G}_0\rangle.
\end{equation}
Note, that the density $\rho_\sigma$ is normalized to $N_\sigma$. In the upper panel of Fig. \ref{Fig2} we show density profiles for $N_\uparrow=3$ and $N_\downarrow=1$ obtained with the exact diagonalisation approach for different $g$. In this way we reconstruct the corresponding result of \cite{ZinnerNJP} by using an exact diagonalisation method which is accurate near $g=0$ for weak interaction. 

The situation is qualitatively different when a difference of mass between flavours is introduced. As an illustration, we repeat all the calculations for $\alpha=40/6$ (lower panel in Fig. \ref{Fig2}). Obviously, in the limit of vanishing interaction the only difference comes from the fact that the heavier impurity is better localised in the middle of the trap. However, for strong enough interactions the separation is not present in the bulk subsystem, but a significant separation of the lighter impurity appears. Note that in this case, the separation of the lighter flavour means that the impurity is in the ''Schr\"odinger cat''-like state, i.e. the probability of finding the particle is non-zero in the two, spatially separated regions of space. It was reported recently that this sort of split state can also be seen for impurities in a bosonic bath \cite{Dehkharghani.QuantumImpurity}. Also a quite similar effect of the phase separation driven by the mass difference was studied recently for spin-1/2 fermions in a uniform trap \cite{JasonHo}.

The mechanism of separation induced by the mass imbalance is fundamentally different form  separation driven by interactions. For example, for equal masses $\alpha=1$, the separation is present in the ground-state only for different number of particles ($N_\uparrow\neq N_\downarrow$) and it appears always in the flavour with larger $N_{\sigma}$. It can be viewed as the consequence of the Pauli exclusion principle and the flavour-exchange symmetry described above. Interactions favour such configurations in which different fermions do not overlap, and as a  consequence particles are promoted to higher single-particle levels respecting the Pauli principle for fermions with the same spin. At the same time, the ground-state of the system has to remain in the {\em symmetric} configuration under exchange of two particles with opposite spins (this is related to the antiferromagnetic structure of the ground state, see \cite{Volosniev2014}). These two requirements force the system to excite favours with larger $N_{\sigma}$ to higher states than the particles of the minority. In consequence, the probability of finding a minority fermion in the centre of the trap decreases slower than for majority fermions. This is the reason why this kind of separation requires an imbalanced particle number. It is worth noticing that this argumentation is valid only for the ground-state of the system which in the limit of infinite interactions becomes degenerate with states of other exchange-symmetries. Those other states may not manifest separation in any of the flavours \cite{Volosniev2014}.

\begin{figure}
 \centering
 \includegraphics[scale=1.2]{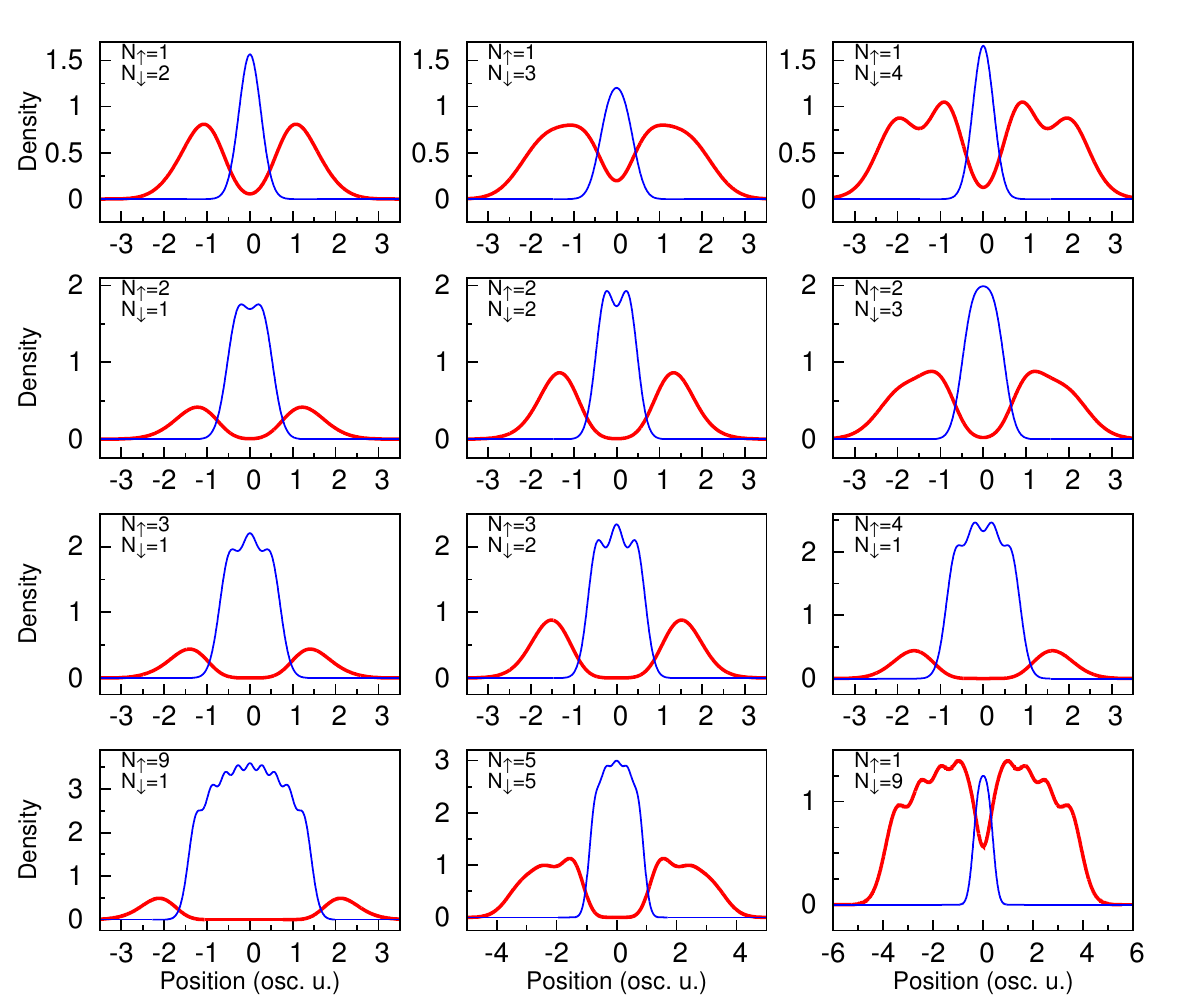}
 \caption{Single-particle densities $\rho_\uparrow(x)$ (thin blue line, heavier flavour) and $\rho_\downarrow(x)$ (thick red line, lighter flavour) calculated in the ground-state of the system for different numbers of fermions with $\alpha=40/6$ and strong interaction $g=10$. In contrast to the previous results obtained for $\alpha=1$, in this case the separation always occurs in the lighter species, independently on the distribution of fermions between flavours. In particular, the separation is present also for an equal number of fermions $N_\uparrow=N_\downarrow$.  \label{fig3} }
\end{figure}

For $\alpha\neq 1$, the separation has a different origin. In this case, the single-particle states $\phi_{i\sigma}$ are not symmetric under flavour-exchange and as a consequence, the interaction energies between the two particles occupying different levels of the harmonic oscillator are not symmetric either. This fact leads directly to the separation in the flavour with smaller mass. To explain this mechanism, let us concentrate on the case $N_\uparrow=3$ and $N_\downarrow=1$.  In the limit of vanishing interactions, the ground-state of the system has the form
$|\mathtt{210;0}\rangle \equiv 1/\sqrt{6}\,\, \hat{a}^\dagger_{2\uparrow}\hat{a}^\dagger_{1\uparrow}\hat{a}^\dagger_{0\uparrow}\hat{a}^\dagger_{0\downarrow}|\mathtt{vac}\rangle$. When interactions are present then particles of different spins try to avoid each other to minimize the interaction energy and as a consequence they change their Fock-space configuration. The interaction energy can be minimized by exciting particles to higher single-particle states. In the model studied here, the energy cost of such an excitation does not depend on the flavour. However, due to the violation of the flavour-exchange symmetry in the interaction sector, the energy gain is larger when lighter particles (i.e. particles with a larger spread of the wavefunction) are promoted. In consequence, the Fock states of the form $|\mathtt{210;k}\rangle \equiv 1/\sqrt{6}\,\, \hat{a}^\dagger_{2\uparrow}\hat{a}^\dagger_{1\uparrow}\hat{a}^\dagger_{0\uparrow}\hat{a}^\dagger_{k\downarrow}|\mathtt{vac}\rangle$ with $k>0$ start to contribute and dominate the exact many-body ground-state of the system. Direct inspection of the ground-state obtained from the exact diagonalisation method proves indeed that this simple heuristic argumentation is correct. For example, for $g=10$ and $\alpha=40/6$ the projection of the true ground state $|\mathtt{G}_0\rangle$ onto the non-interacting ground-state is $|\langle\mathtt{210;0}|\mathtt{G}_0\rangle|^2<10^{-4}$. At the same time, the cumulative probability of finding all three heavy particles in the lowest states, i.e. $\sum_k|\langle\mathtt{210;}k|\mathtt{G}_0\rangle|^2\approx 0.9$. This means that the energy of the ground-state is minimised mainly due to the mobility of the lighter particle, i.e. the particle with the larger spreading of single-particle states.  

We have checked that the mechanism described above is very general and does not depend on the number of fermions of different flavours. In Fig. \ref{fig3} we present the single-particle densities for different number of fermions $N_\uparrow$ and $N_\downarrow$ obtained in the strong interaction regime $g=10$. In all plots, the mass ratio is  $\alpha=40/6$. It is seen that the separation induced by a mass difference always occurs in the lighter component (red thick line). Our results performed for all configurations with $N_\uparrow+N_\downarrow\le 10$ confirm that the lighter component is always divided into two domains regardless of the number of fermions.

It is worth noticing that for $N_\uparrow=1$ and $N_\downarrow=2$ our results obtained in the strong interaction limit are consistent with the density profiles obtained in \cite{loft2014variational} where a variational approach was used for this three-body problem.

\section{Non-zero temperatures}
\begin{figure}
 \centering
 \includegraphics[scale=1.2]{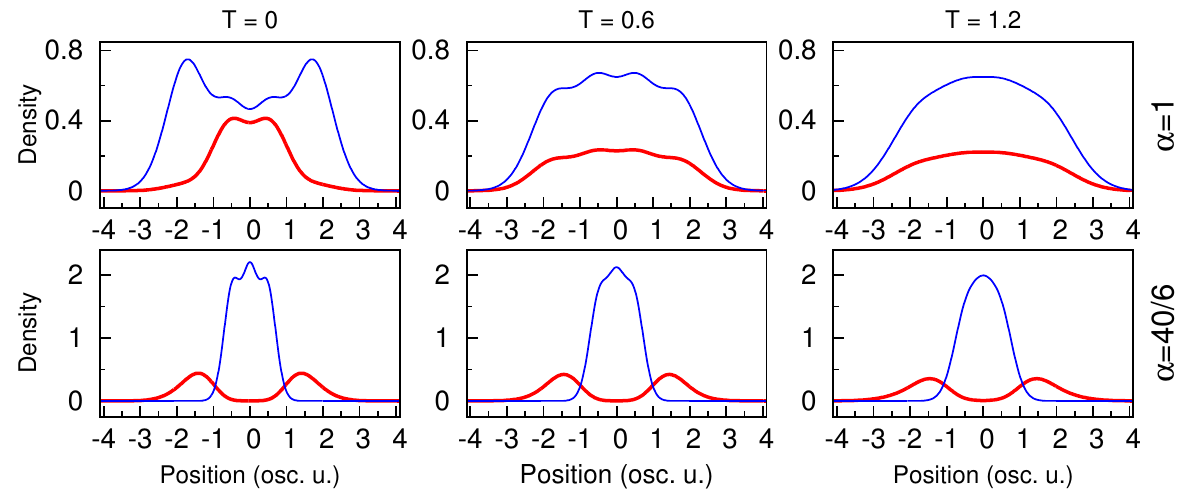}
 \caption{Single-particle densities $\rho_\uparrow(x)$ (thin blue line, heavier flavour) and $\rho_\downarrow(x)$ (thick red line, lighter flavour) calculated in the thermal state $\hat\rho_{\mathtt{T}}$ with $N_\uparrow=3$ and $N_\downarrow=1$ for $g=10$ as a function of temperature $T$. For $\alpha=1$ the phase separation induced by the imbalanced number of fermions is instantly destroyed at non-zero temperature. In contrast, for $\alpha=40/6$, when the phase separation is driven by a mass difference between fermions, the single-particle density of lighter flavour is almost unchanged and it is very robust to the mixing with higher many-body states. \label{fig4} }
\end{figure}

The method used allows us to study the properties of the system also at non-zero temperatures, i.e. in a mixed state of the form $\hat\rho_{\mathtt{T}} ={\cal Z}^{-1} \sum_i p_i |\mathtt{G}_i\rangle\langle\mathtt{G}_i|$, where $p_i= \mathrm{exp}(-{\cal E}_i /k_B\,T)$ and ${\cal Z}=\sum_i p_i$ is the partition function. A discussion of whether the one-dimensional system described by the Hamiltonian (\ref{Ham1}) reaches the thermal state $\hat\rho_{\mathtt{T}}$ when interacting with the thermal bath is beyond the scope of this article. However, our aim is to discuss the influence of the interaction with an environment on the phenomenon of phase separation. From this point of view, the temperature $T$ can be viewed as an effective parameter which controls strength of the interaction with an external environment.

After defining the mixed state $\hat\rho_\mathtt{T}$, the whole analysis can be performed in analogy to the zero temperature case provided that the single-particle densities $\rho_\sigma(x)$ are calculated in the state $\hat{\rho}_{\mathtt{T}}$, i.e.
\begin{equation} \label{eq:ThermalDensity}
  \rho_{\sigma}(x,T) = \mathrm{Tr} \left[\hat{\rho}_{\mathtt{T}}\,\hat\Psi_\sigma^\dagger(x)\hat\Psi_\sigma(x)\right].
\end{equation}
In practice, we cut  the number of exact many-body eigenstates at the number $M$ (defined in Sec. 3) which is temperature dependent. For a given temperature, this number is determined from the condition that the final single-particle densities do not change significantly when $M$ is increased. It is natural that in the limit $T\rightarrow 0$ we reproduce the previous results obtained for $\rho_\sigma(x)$ defined by eq. (\ref{DensProf}). 
 
In Fig. \ref{fig4} we visualize the influence of non-zero temperatures on the phase separation for a number of particles equal to $N_\uparrow=3$, $N_\downarrow=1$ and strong interactions $g=10$. As seen in the upper panel for $\alpha=1$, the phase separation is destroyed almost immediately when the higher eigenstates of the Hamiltonian contribute. Density profiles are becoming smoother and spatial density fluctuations vanish. In contrast, for $\alpha=40/6$, the separation induced by a large mass imbalance is almost insensitive to the temperature, i.e. the lighter particle remains in the Schr\"odinger cat-like state. The only visible consequence of higher temperature is a smoothing of the bulk flavour density. This result confirms that the mechanism of the separation induced by a mass imbalance between flavours has a completely different origin than the separation induced by interactions for a system with flavour-exchange symmetry. 

\section{Conclusions}
We have studied the properties of a two-component mixture of fermions in a one-dimensional harmonic trap for different mass ratios between atoms. We show that the mass difference induces a phase separation in the lighter flavour independently of the distribution of fermions between the flavours. The mechanism observed is general and is forced by the interaction energy cost when particles are excited to higher states of the harmonic oscillator. Finally, we show that in contrast to the standard separation induced by an imbalanced  number of fermions, the separation driven by the mass difference is robust to non-zero temperatures. We believe that our predictions can be confirmed in upcoming experiments with ultra-cold mixtures of fermions. 
   
\section{Acknowledgements}  
The authors thank P. Deuar for his fruitful comments and suggestions.
This work was supported by the (Polish) Ministry of Sciences and Higher Education, Iuventus Plus 2015-2017 Grant No. 0440/IP3/2015/73 (D.P. and T.S.) and by the (Polish) National Science Center Grant No. DEC-2012/04/A/ST2/00090 (M.G.).

\end{document}